# Optimization of self-interstitial clusters in 3C-SiC with Genetic Algorithm

Hyunseok Ko[a], Amy Kaczmarowski[a,b], Izabela Szlufarska[a], Dane Morgan[a, *]

[a] *Department of Materials science and Engineering, University of Wisconsin, Madison, WI 53706, USA*

[b] *Sandia National Laboratories, Albuquerque, NM 87123, USA*

## ABSTRACT

Under irradiation, SiC develops damage commonly referred to as black spot defects, which are speculated to be self-interstitial atom clusters. To understand the evolution of these defect clusters and their impacts (e.g., through radiation induced swelling) on the performance of SiC in nuclear applications, it is important to identify the cluster composition, structure, and shape. In this work the genetic algorithm code StructOpt was utilized to identify groundstate cluster structures in 3C-SiC. The genetic algorithm was used to explore clusters of up to ~30 interstitials of C-only, Si-only, and Si-C mixtures embedded in the SiC lattice. We performed the structure search using Hamiltonians from both density functional theory and empirical potentials. The thermodynamic stability of clusters was investigated in terms of their composition (with a focus on Si-only, C-only, and stoichiometric) and shape (spherical vs. planar), as a function of the cluster size ($n$). Our results suggest that large Si-only clusters are likely unstable, and clusters are predominantly C-only for $n \leq 10$ and stoichiometric for $n > 10$. The results imply that there is an evolution of the shape of the most stable clusters, where small clusters are stable in more spherical geometries while larger clusters are stable in more planar configurations. We also provide an estimated energy vs. size relationship, $E(n)$, for use in future analysis.

*Corresponding Author. Tel:+1-608-265-5879 ; Email: ddmorgan@wisc.edu (Dane Morgan).

# 1. Introduction

Silicon carbide (SiC) is a promising nuclear material owing to its superior thermo-mechanical properties [1,2]. The proposed applications include a replacement for Zircaloy cladding in light water reactors, and the primary structural and fission product barrier layer in Tristructural-Isotropic (TRISO) coated fuel particles [3]. With the increasing interest in SiC, a substantial effort in the last few decades has been devoted to elucidating phenomena related to radiation damage and the effect of radiation on SiC properties [3]. A fundamental understanding of radiation damage in SiC, particularly of the defect evolution, is needed for development of predictive models of this material in nuclear applications. Under thermodynamic equilibrium conditions, the concentrations of point defects in SiC are extremely low due to their high formation energies [4,5]. However, during neutron irradiations or ion implantations, lattice defects are created in greater amounts than their equilibrium concentrations. The accumulation of irradiation-induced lattice damage can lead to unwanted microstructural changes. In particular, radiation induces defects such as dislocations, Frank dislocation loops, voids, and so-called black spot defects (BSDs), which name is due to their appearance as black spots in bright field transmission electron microscope (TEM) images [6]. Over time radiation induced defects can be responsible for radiation effects such as amorphization [7,8], swelling [9], and degradation of the thermo-mechanical [10] properties of SiC.

In both neutron and ion irradiated SiC under irradiation conditions of low dpa (<10) and temperature below ~1400 °C uniformly dispersed BSDs are observed from bright field image TEM [6] or scanning TEM (STEM) [11]. The BSDs are typically observed with a diameter of 1-5 nm with a number density in the orders of $10^{22}$-$10^{24}$ m$^{-3}$ [6,12,13]. The BSDs, which are speculated to be self-interstitial atom (SIA) clusters, are quite stable, and persist long after irradiation. Such microstructural changes can cause degradation of the material's properties, such as radiation-induced swelling [9,14,15]. They can also become nuclei for growth of larger defect clusters, and therefore understanding structure and energetics of BSD is a critical part of any radiation damage model. At present this understanding is limited. Most of the previous studies [16-19] focused on C interstitials, likely because C Frenkel pairs are produced in higher quantities than Si Frenkel pairs during irradiation [20,21]. Additionally, C interstitials have a lower kinetic barrier to diffusion and therefore are likely to cluster faster (migration barriers are 0.74 eV [5]). One example of studies of C clusters in SiC was reported by Mattausch, *et al.* [18], who identified a number of stable structures for small C interstitial clusters (1-4 interstitials) from first principle calculations. Jiang *et al.* [17] made further progress in identifying groundstate (GS) and meta-stable structures of small C clusters of size 1-6 interstitials. Jiang *et al.* [17] used both molecular dynamics and Monte Carlo basin-hopping simulations to find GSs, typically starting with interatomic potentials (they used the EDIP potential [22]) and then calculating energies of the most promising configurations with *ab initio* methods. On the other hand, there are only few studies available on clusters consisting of Si or mixed compositions. The Si interstitials, concentration of which is significant under irradiation, can be quite mobile even at low temperatures [23,24]. Liao and Roma [25] showed that Si tri-interstitials have strong binding energies and suggested that the clustering of Si interstitials is also energetically favorable. For larger clusters of a few hundred interstitials, Watanabe, Morishita and Kohyama [26] used empirical potential to investigate the size dependence of the formation energy for planar SIA-clusters with Si-only, C-only, and stoichiometric compositions. The stoichiometric SIA was predicted to be the most stable composition for clusters of $n = \sim100 – 300$ in Ref. [26]. From studies available at the time of this writing, it still remains uncertain what kind of clusters, e.g. their composition and shape, are stable as a function of cluster size ($n$). In particular, if small clusters tend to be C-only or C-rich, when and how they evolve to a stoichiometric composition [26] remains uncertain. Also, it remains unexplained when and how small clusters without planar characteristics evolve to the stable planar structures, which are expected for large clusters [15].

The goals of the present study are (i) to systematically investigate the structures of SIA clusters ($n$ up to 30-interstitials) in 3C-SiC, and (ii) to determine the thermodynamic stabilities of these clusters. In this study, an automated search was performed across the composition of clusters for the first time. Both density functional theory (DFT) and interatomic potentials were used depending on the target cluster size, as the DFT is expected to be more accurate than potentials but is too slow to use for the larger clusters. In particular, DFT was applied to all compositions for $n \leq 4$, and an interatomic potential was employed to model three compositions (Si-only, C-only, and stoichiometry (Si=C)) for $n > 4$. After identifying the most stable states at a given composition and size, we will discuss the stability of the clusters in terms of their binding energies and as well as the compositions and shapes of stable clusters as a function of their size.

Throughout the paper, we notate the composition of a $n$ interstitial cluster, consisting of $n_C$-carbon and $n_{Si}$-silicon interstitials, as $Si_{n_{Si}}C_{n_C}$. Furthermore, we will adopt the following conventions for naming the structures of clusters. For a structural unit of a cluster with $n_{Si}$ silicon atoms, $n_C$ carbon atoms, we will denote the subunit with the symbol Si$n_{Si}$C$n_C$. If the subunit can be identified with a lattice site, then we add a subscript X and denote the structural unit by (Si$n_{Si}$C$n_C$)$_X$. If the structural unit can be further identified by a simple configuration description, Y, then we denote the structural unit by (Si$n_{Si}$C$n_C$)$_{X,Y}$. Finally, we denote the complete structure of multiple structural units by joining the structural units with a dash, and including the number of times each structural unit is repeated as a coefficient of that structural unit, e.g., 2(Si$n_{Si}$C$n_C$)$_{X,Y}$ - (Si$n'_{Si}$C$n'_C$)$_{X',Y'}$. Values of X are Si or C sites, Si-C nearest-neighbor bond center site (BC), a tetragonally C coordinated site (TC), and a tetragonally Si coordinated site (TSi). Value of Y are a triangular arrangement of interstitials (triangle), and a split-interstitial (sp) or the direction of split interstitial (e.g., <100>). Note that a single lattice site X can have multiple species associated with it if they are all equally close. Some examples of this notation are: (i) a C-C split interstitial is denoted (C$_2$)$_{C,<100>}$ as this is two C atoms associated with one C site and oriented along <100> and (ii) the three Si interstitial cluster groundstate is denoted as 2(Si$_1$)$_{BC}$ –(C$_1$)$_{BC}$–Si$_C$ as this structure contains 2 Si interstitial atoms at distinct bond centers, one C interstitial at a third bond center, and one Si on a C lattice site (a Si antisite defect).

## 2. Methodology

### 2.1. Genetic Algorithm: GS search method

The stable structures of multi-atom defect clusters embedded in a crystal are challenging to determine due to there being many possible metastable local equilibria. In this study, the defect structures will be determined with a real-space genetic algorithm (GA) approach. The genetic algorithm is a heuristic search that mimics the biological evolution of a population of distinct individuals, and the process of natural selection. This heuristic search can effectively explore complex energy landscape through various crossover and mutations, and eventually outputs candidate solutions to optimization of a given function, which would be the lowest energy in this case. A recent study by Kaczmarowski *et al.* [27] developed a formalism for applying GA for defect clusters in crystals. The authors demonstrated that this method is an efficient automated way to successfully predict small interstitial clusters embedded in bulk materials, with specific examples of cubic SiC, BCC Fe, and BCC Fe–Cr random alloys. This formalism has been released as an open source code called StructOpt [27], which is part of the MAterial Simulation Toolkit (MAST) [28] to help manage the StructOpt workflow (Both StructOpt and MAST are available at https://pypi.python.org/pypi/MAST). Throughout this study MAST and StructOpt were used. The general descriptions of inputs for StructOpt can be found in Ref. [27], and the specific choices of input parameters are summarized in the Supplementary Information (S.I.) Section A.

The parameters used in this work are summarized in S.I.. The GA manipulates the real-space coordinate positions and types of the atoms both in and near the defect (the cluster regions) over several generations to drive the system towards lower energy structures. The simulation cell for GA calculation is divided into three regions (Set 1, 2 and 3) as shown in Figure 1 – Set 1 defines the volume where initial defects are located, Set 2 is a bulk region surrounding the defect, and Set 3 contains the remainder of the bulk material. The cluster regions (Set 1 and 2) are isolated from the bulk structure (Set 3) before any manipulation of the cluster regions occurs, so the host bulk lattice is left intact as the cluster evolves. In each GA calculation, a fixed number of interstitials, with a constant Si/C ratio are introduced into the bulk. The GA outputs the lowest energy structure at given local composition in the Set 1 and 2. When *n* interstitials are introduced, they are distributed within the volume of Set 1. Further details on the defect optimization scheme can be found in S.I. Section A.

### 2.2. Energy evaluation methods in genetic algorithm

The *ab initio* DFT calculations were performed with the Vienna Ab-Initio Simulation Package (VASP) [29-32], and the projector-augmented plane-wave (PAW) method [33,34] was used for the atomic potentials. The electronic state of the valence electrons in the potentials are $3s^2\ 3p^2$ for Si, and $2s^2\ sp^2$ for C. Due to the complex structures being explored we used the hardest potentials available, labeled Si_h and C_h in the VASP files, with ENMAX cutoffs of 380 and 700 eV, respectively. The exchange-correlation was treated in the Generalized Gradient Approximation (GGA), as parameterized by Perdew, Burke, and Ernzerhof (PBE) [35]. For GA optimization, a single Γ-point was used to sample the reciprocal space, and the energy cut-off was set at 700eV, which is the largest ENMAX of the pseudopotentials used in this work. At each GA step, the ions are relaxed to a local minimum. The convergences for the ion relaxation and electron self-consistency cycle were set to $10^{-3}$ and $10^{-4}$ eV, respectively. The error associated with the *k*-point mesh was about 4 meV/atom for a 64-atom cell (2×2×2 supercell of the conventional cell), as compared to the 3×3×3 *k*-point mesh. The final output structures from each completed GA run are finally relaxed with the 3×3×3 *k*-point mesh, to improve the accuracy of the calculations.

For interatomic potential optimization, the Large-scale Atomic/Molecular Massively Parallel Simulator (LAMMPS) [36] package was utilized. For SiC, many parameterizations of empirical potentials have been developed in the literature [22,37-42] and the best choice of the empirical potential is problem-dependent. In this work the Gao–Weber (G-W) potential [16] was used, which was developed to improve the description of SIA in cubic SiC by modifying the Brenner potential [43]. The G-W potential was demonstrated to provide a reasonably good description of the structures and energies of self-interstitials when compared with DFT results [16].

A 64-atom structure (2×2×2 supercell of the conventional cell) was used for all DFT calculations, and a 1728-atom structure (6×6×6 supercell of the conventional cell) was used for G-W potential calculations. For the DFT calculation, where the finite size supercell can lead to inaccuracies in the desired estimate of the energy of an isolate defect, a larger cell of 216-atoms was also used. The 216-atom cell was explored for the $Si_4$ and $Si_3C_1$ clusters, as these are the largest clusters we studied with DFT and finite size effects are expected to be most important. The GS for $Si_4$ was found to be different between the 64- vs. 216-atom cells. The GS of $Si_4$ in the 216-atom cell was 0.12 eV/Int. (hereafter, energies are compared in eV per interstitial, denoted eV/Int.) more stable than the GS in the 64-atom cell, and so the 216-atom cell results are used in the rest of the text. The GS for the $Si_3C_1$ cluster was found to be the same for the 64- and 216-atom cells, with energy difference less than 0.03 eV/Int., suggesting that the 64-atom cell was appropriate for identifying the structure this cluster and other small clusters. In GA calculations with both DFT and G-W, the structures were relaxed at 0K with a fixed cell shape, but they were allowed to change volume. Periodic boundary conditions are applied in all spatial directions. It should be noted that charged supercells or explicitly charged defects were not considered in this study.

For the formation energy ($E_F$) of defects, we use the following expression:

Equation 1: $E_F = E_{def} - E_{undef} + \Sigma_i \Delta n_i \mu_i$

where $E_{def}$ and $E_{undef}$ are energies of the defected and the undefected cell, $\Delta n_i$ is the change in the number of the atomic species $i$ (Si or C) in the defected cell from the number of same species in the undefected cell, and $\mu_i$ is the chemical potential of the species $i$ in SiC relative to its reference state (diamond lattice for Si and graphite for C). The formation energy of SiC per formula unit with DFT and G-W potential are calculated as -0.45 and -4.69 eV, respectively. As G-W potential was primarily designed to provide satisfactory descriptions for point defects, the bulk energy for Si and C are rather far from an agreement with DFT (Table B.1. in Supplementary Information) and therefore there is a large discrepancy in the SiC formation energy. The chemical potential values for the different external environments are typically taken for three cases: excess Si (Si-rich), excess C (C-rich), and stoichiometric (halfway between two extremes, or $\mu_{Si} = \mu_C$). Throughout this paper, the stoichiometric chemical potentials will be used to calculate the formation energies. In the S.I. Section B, we provide $\mu_i$ values for the three external chemical potential states, for cases where the energy values need to be referenced to other states.

To evaluate the stability of SIA clusters, we introduce the dissociation energy ($E_D$), which is the energy required to fully dissociate $n$-SIA cluster ($I_n$) into $n$ non-interacting mono-interstitials ($I_1$). $E_D$ is defined as

Equation 2:    $E_D = [n \cdot E_F(I_1) - E_F(I_n)] / n$

where $E_F(I_1)$ is the formation energy of a mono-interstitial in its GS configuration. The values of $E_D$ are a measure of how stable SIAs are as a cluster (positive representing a cluster that is more stable than the isolated interstitials).

We found that the mono-interstitial GSs from DFT for C and Si interstitials, respectively, were tilted split C-C along <100> direction, which has the energy of 4.81 eV, and split Si-Si along <110> direction, which has the energy of 10.47 eV (Fig. 1a). These findings are in an agreement with previous studies [17,24]. We found that the mono-interstitial GSs from interatomic potential optimization were $(C_2)_{C,<100>}$ with an energy of 3.04 eV for a C interstitials, and $(Si_1)_{TC}$ with an energy of 3.36 eV for a Si interstitial. These findings also agree well with the previous study by Gao and Weber [16]. The comparison of energies and structures between DFT and G-W potential will be further discussed in later Sec. 3.2.1.

## 3. Results

### 3.1. DFT optimized SIA clusters of $n$=1-4

#### 3.1.1. Structures and energies of clusters

The structures and configurations of $n$-SIA clusters optimized by GA-DFT calculations are summarized in Table 1. Coordinates of the clusters (all composition for $n \leq 4$, and Si=C for $n > 4$) found in this study are reported in Supplementary Information. Here we will first discuss each GS structures, and if available, make a comparison to previously reported structures.

Table 1. The formation and dissociation energies ($E_D$) of DFT optimized $n$-atom SIA clusters with $n_C$-carbon and $n_{Si}$-silicon SIAs. $E_D$ is referenced to mono-interstitial DFT energies and it is positive when the cluster is more stable than the isolated point defects that comprise the cluster. Energies of the most stable structures with highest $E_D$ at the given local composition are presented. The structural information for GSs can be found in Figure 2-4.

| $n$ | $n_C$ | $n_{Si}$ | Structure | $E_F$ (eV/Int.) | $E_D$ (eV/Int.) This work | $E_D$ (eV/Int.) Ref. |
|---|---|---|---|---|---|---|
| 1 | 0 | 1 | $(Si_2)_{Si, <110>}$ | 10.47 | - | - |
| 1 | 1 | 0 | $(C_2)_{C, <100>}$ | 4.81 | - | - |
| 2 | 0 | 2 | $(Si_3)_{Si, triangle}$ | 9.62 | 0.85 | 0.82 [25] |

| | 1 | 1 | $(C_1Si_1)_{TC}$ | 7.32 | 0.30 | - |
| | 2 | 0 | $2(C_1)_{BC}$ | 2.39 | 2.35 | 2.58 [17] (2.35[b]) |
| 3 | 0 | 3 | $2(Si_1)_{BC'}-(C_1)_{BC'}-Si_C$ | 7.33 | 3.13 | - |
| | 1 | 2 | $2(Si_1)_{BC'}-(C_1)_{BC'}$ | 7.20 | 2.90 | - |
| | 2 | 1 | $(Si_1)_{BC'}-2(C_1)_{BC'}$ | 4.81 | 2.57 | - |
| | 3 | 0 | $3(C_1)_{BC'}$ | 1.87 | 2.90 | 3.17 [17] (2.90[b]) |
| 4 | 0 | 4 | $(Si_1)_{TC'}-2(Si_1)_{BC'}-(C_1)_{BC}-Si_C$ | 8.87 | 1.60 | - |
| | 1 | 3 | $(Si_3)_{Si,triangle}-(Si_1C_1)_{TC}$ | 6.52 | 2.53 | - |
| | 2 | 2 | $2(Si_1)_{BC'}-(C_1)_{BC'}-(Si_1C_1)_{Si}$ | 5.13 | 2.48 | - |
| | 3 | 1 | $(Si_1C_1)_{BC'}-2(Si_1C_1)_{Si}$ | 3.41 | 2.80 | - |
| | 4 | 0 | $2(C_2)_{BC}$ | 1.66 | 3.15 | 3.09[a] [17] (2.90[b]) |

[a] Reference structure is different from the GS structure identified in this study
[b] Reference structure relaxed with input parameters used in this study

In Figure 2 (c-e), the GSs of $n=2$ clusters are summarized. The GS of $Si_2$ had an isosceles triangle structure sharing the base bond center on a Si sublattice site (Figure 2(c)). The bond length of the base bond (shorter bond) and the isosceles bonds are 2.11 and 2.27 Å, respectively. The $E_D$ of this isosceles triangle is found to be 0.85 eV/Int. A similar structure was reported in DFT study by Liao *et al.* [25] who identified it using both relaxation and saddle point search. However, there are differences between our GS of $Si_2$ and that of Liao *et al.*, which will be discussed below. The GS of $C_2$ was found in a $(C_2)_{BC}$ structure where the center of bond between two C interstitials occupy the bond-center (BC) positions of Si and C sublattices (Figure 2(e)), which is in an agreement with the previous studies [17-19,44]. The $E_D$ of this cluster was 2.4 eV/Int., which is also in an agreement with previous studies. The $Si_1C_1$ composition cluster has not been studied previously as far as we are aware. It has its bond center at the TC site with $E_D$ of 0.3 eV. This relatively low dissociation energy suggests that the interstitials can easily unbind at non-zero temperature.

Overall, the $n=1-2$ GS structures identified in this work were consistent with previous DFT studies when data for comparisons were available [17-19,25,44,45]. However, there are a few differences between our $Si_2$ GS structure and the GS in Liao *et al.* [25], such as bond lengths and orientation of the isosceles triangle. The structure from Liao *et al.* had bond lengths of 2.08 and 2.19 Å, for the base bond (shorter bond) and the isosceles bonds, respectively. Also the structure reported in Ref. [25] had a center of the base bond on the Si sublattice site, which is inconsistent with the GA structure found in our work. To determine if the disagreement between the two structures was due to differences in computational methods, we relaxed both structures by applying the same set of computational parameters. First, the structures from GA and Liao *et al.* [25] are relaxed with our set of parameters, then relaxed with the parameters given by Liao *et al.*. In both cases the GA structure was found to be significantly more stable. Using parameters described earlier in our current paper, the formation energies (stoichiometric chemical potential condition) of structures from GA and Liao *et al.* were calculated to be 7.94 and 8.91 eV/cluster, respectively. In general, the di-Si interstitial appears to have many metastable states with an isosceles triangle like structures (many were observed during GA optimization) making identification of the GS particularly challenging.

In Figure 3, the GS structures of $n=3$ clusters are shown. For $Si_3$, the GS has two Si interstitial atoms and one C interstitial occupying bond center between the Si-C bonds, but off from the BC (which sites we denote as BC'), with a Si antisite defect (Figure 3(a)). The interstitials form a nearly equilateral triangle structure approximately parallel to the {111} plane. It is perhaps surprising that the structure initialized with three Si SIA forms a GS with two Si SIAs, one C SIA, and a Si antisite. However, as mentioned in Sec. 2.1, we have allowed sublattice atoms to form defects (e.g., antisites) if they are stable for a given local composition, even though the structures are not precisely interstitial clusters. This result shows that in the presence of other Si SIAs, the exchange of $Si_1 \rightarrow Si_C + C_1$ can become energetically favorable (note that, for an isolated Si interstitial, this reaction requires 1.7 eV). The structures for mixed compositions, $Si_2C_1$ and $Si_1C_2$, have interstitials between the Si-C bonds forming isosceles triangles without any antisites (Figure 3(b-c)), again approximately parallel to the {111} plane. The GS of C-only ($C_3$) SIA again forms an equilateral triangle structure with three C interstitials in $(C_1)_{BC}$ (Figure 3(d)), which structure is consistent with the GS identified by Jiang, *et al.* [17]. It is Interesting to note that the configuration of a triangle of BC or near BC (BC') interstitials, approximately parallel to the {111} plane, was found for all $n=3$ clusters, as it can be seen in Figure 3.

The GSs of tetra-SIA clusters ($n=4$) were found to have more complicated configurations, which are somewhat difficult to describe due to the large displacements of atoms around the SIAs. The Si-only cluster, $Si_4$ forms $(Si_1)_{TC}–2(Si_1)_{BC}–(C_1)_{BC}–Si_C$ (Figure 4(a)), and like $Si_3$ cluster, involved the formation of an antisite. Antisite formation in tri- and tetra- Si Int. clusters indicate that $Si_1 \rightarrow Si_C + C_1$ can become energetically favorable for Si-only clusters. The structure for $Si_3C_1$ (Figure 4(a)) can be seen as two smaller sub-clusters: (i) two Si interstitials form a non-isosceles triangle on Si sublattice sites, which is similar to the $Si_2$ structure in Figure 2(c) and (ii) a Si-C di-interstitial forms a bond center at the TC site, which is also similar to the isolated Si-C di-interstitial structure predicted in Figure 2 (d). These sub-clusters together are 8 eV more stable than isolated sub-clusters, so they are strongly bound together and should be considered as one cluster. The GS structure of $Si_2C_2$ composition in Figure 4(c) was found in the form of a tilted triangle structure, containing two Si and one C at BC positions, and a nearby Si-C dumbbell on the Si sublattice. For $Si_1C_3$, the chain of interstitial atoms and few sublattice atoms form a screw configuration with an axis along the <100> direction as shown in Figure 4(d). The GS structure of $C_4$ in Figure 4(e) can also be described as two sub-clusters, which are two di-interstitial clusters in $(C_2)_{BC}$ configuration with different orientations. The $2(C_2)_{BC}$ structure for $C_4$ reported in this study has $E_D$ of 3.1 eV/Int., which is more stable than our calculated $E_D$ of 2.9 eV/Int.. for the previously reported $4(C_2)_{sp}$ structure proposed as GS by Jiang *et al.* [17].

### 3.1.2. Thermodynamic stability of clusters

The thermodynamic stability is an important property for predicting the behavior of generated interstitials (e.g., whether they cluster or remain isolated from each other and diffuse to defect sinks) and to provide further insights into the evolution of these clusters under irradiation and thermal aging. In Figure 5 and Table 1, the dissociation energies ($E_D$) are shown for clusters of $n \leq 4$. For all clusters identified here, the dissociation energies were positive, indicating that the clusters of SIAs are always more stable than the corresponding isolated interstitials. For $n=2–4$, the C-only clusters are found with highest $E_D$ at a given $n$, except for $n=3$. This suggests that C clusters might be easier to form than Si clusters, a point that will be discussed further in Sec. 4. For $n=3$, the Si-only cluster has the highest $E_D$, but this $Si_3$ structure involves a Si antisite, making it somewhat unusual. This issue will be discussed in Sec. 4 as well.

To better assess cluster thermodynamics in the context of their tendency to grow and their long-term stability, we consider the following binding energies:

Equation 3 a.  $E_B = [E_F(I_{n-1}) + E_F(I_1)] - E_F(I_n)$

Equation 3 b.  $E_B' = [E_F(I_{n-2}) + E_F(I_1) + E_F(I_1')] - E_F(I_n)$

Equation 3 c.  $E_B'' = [E_F(I_{n-2}) + E_F(I_2)] - E_F(I_n)$

Here, $E_B$ is the energy to dissolve one interstitial from a cluster (Equation 3 a), $E_B'$ is the energy to dissolve two interstitials from a cluster (Equation 3 b), and $E_B''$ is the energy to dissolve a sub-cluster of 2 interstitials from a cluster (Equation 3 c). The calculated values of $E_B$, $E_B'$, and $E_B''$ are summarized in Table 2. Positive values of these quantities imply that the binding of an interstitial or a smaller cluster to another cluster is favored. Thus a cluster with positive binding energies is expected to grow under irradiation in the presence of a supersaturation of interstitials or mobile small clusters [46]. All compositions of SIA clusters reported in this study, except $Si_4$, were found to have positive binding energies. The Si-only cluster with $n=4$ has $E_B < 0$, which means that the unbinding of Si interstitial from the $Si_4$ cluster is thermodynamically favored. In Sec. 4, the Si-only cluster and its lack of stability will be discussed further.

Table 2. The binding energy of DFT-optimized SIA clusters. The binding energies ($E_B$, $E_B'$, and $E_B''$) are defined as the change in the enthalpy when a cluster unbinds interstitial(s) or sub-clusters. Specific expressions for these energies are shown as Equation 3.

| $n$ | $n_C$ | $n_{Si}$ | $E_B$ (eV/Int.) | | $E_B'$ (eV/Int.) | | | $E_B''$ (eV/Int.) | | |
|---|---|---|---|---|---|---|---|---|---|---|
| | Species unbinding | | Si | C | Si | Si, C | C | 2Si | Si+C | 2C |
| 1 | 0 | 1 | - | - | - | - | - | - | - | - |
| | 1 | 0 | - | - | - | - | - | - | - | - |
| 2 | 0 | 2 | 1.7 | - | - | - | - | - | - | - |
| | 1 | 1 | 0.6 | 0.6 | - | - | - | - | - | - |
| | 2 | 0 | - | 4.7 | - | - | - | - | - | - |

| n | nC | nSi | | | | | | | | |
|---|----|-----|---|---|---|---|---|---|---|---|
| 3 | 0 | 3 | 8.1 | - | - | - | - | 8.1 | - | - |
| 3 | 1 | 2 | 8.4 | 7.2 | - | - | - | 7.2 | 8.4 | - |
| 3 | 2 | 1 | 3 | 7.2 | - | - | - | - | 7.2 | 3 |
| 3 | 3 | 0 | - | 3.9 | - | - | - | - | - | 3.9 |
| 4 | 0 | 4 | **-3.5*** | - | 4.6 | - | - | 2.8 | - | - |
| 4 | 1 | 3 | 1.4 | 0.5 | 9.8 | 8.6 | - | 8.0 | - | - |
| 4 | 2 | 2 | 2.2 | 4.6 | 5.2 | 9.4 | 8.2 | 3.4 | 8.8 | 3.4 |
| 4 | 3 | 1 | 2.5 | 3.4 | - | 6.4 | 6.4 | - | - | 5.8 |
| 4 | 4 | 0 | - | 3.8 | - | - | 7.7 | - | - | 2.9 |

### 3.2. G-W optimized SIA clusters of $n$=1-30

In the Table 3, the GA optimized GS energies for $n = 1-4$ are shown, and compared to the earlier studies that also used the G-W interatomic potential. A good agreement with Gao and Weber [16] and Watanabe *et al.* [47] (where GSs were identified by molecular static-molecular dynamics modeling for $n =1-6$ clusters) demonstrates that our GA search has been successful for the moderately small space search, although GS structures have not provided in Ref. [47] to make the direct comparison of atomic configurations possible. Table 3 shows that in many cases the GA has identified more stable structures with lower values of $E_F$ (e.g., lower by 0.21 eV/Int., for C$_3$). In Table 4, we summarize the binding energies, $E_B$, $E_B{'}$, and $E_B{''}$ (defined in Equation 3 (a-c)) for G-W optimized clusters. All G-W potential optimized clusters showed positive binding energies, implying that the binding of an interstitial or a smaller cluster to another cluster is favored for all the listed sizes and compositions.

Table 3. The formation and dissociation energies of $n$-atom clusters with $n_C$-carbon and $n_{Si}$-silicon SIAs, optimized by Gao-Weber potential. The formation energy calculated by Gao and Weber [16] and by Watanabe *et al.* [47] are listed for comparison. The dissociation energy is referenced to mono-interstitial GS energies.

| $n$ | $n_C$ | $n_{Si}$ | Structure | $E_F$ (eV/Int.) | | | $E_D$ (eV/Int.) |
|---|---|---|---|---|---|---|---|
| | | | | This work | Ref. [16] | Ref. [47][b] | |
| 1 | 0 | 1 | (Si$_1$)$_{TC}$ | 3.36 | 3.43[a] | 3.37 | 0.00 |
| 1 | 0 | 1 | (Si$_2$)$_{Si, <110>}$ | 4.93 | - | - | 1.57 |
| 1 | 1 | 0 | (C$_1$)$_{TSi}$ | 4.32 | 4.32 | - | 1.28 |
| 1 | 1 | 0 | (C$_2$)$_{C, <100>}$ | 3.04 | 3.04 | 3.04 | 0.00 |
| 2 | 0 | 2 | (Si$_2$)$_{TC}$ | 2.39 | - | 2.41 | 0.97 |
| 2 | 1 | 1 | (Si$_1$)$_{TC}$-(C$_1$)$_{TSi}$ | 2.66 | - | 2.67 | 0.54 |
| 2 | 2 | 0 | 2(C$_1$)$_{BC}$ | 2.33 | - | 2.33 | 0.70 |
| 3 | 0 | 3 | (Si$_3$)$_{TC}$ | 2.30 | - | 2.30 | 1.06 |

| | 1 | 2 | $(Si_1)_{TC}-(C_2)_C-(Si_1)_{TC}$ | 2.30 | - | 2.35 | 0.95 |
| | 2 | 1 | $(C_2)_{BC'}-(Si_1)_{TC}$ | 2.25 | - | 2.35 | 0.90 |
| | 3 | 0 | $3(C_1)_{BC'}$ | 1.88 | - | 2.09 | 1.16 |
| | 0 | 4 | $4(Si_1)_{TC}$ | 1.94 | - | 1.94 | 1.42 |
| | 1 | 3 | $(C_2)_{C, <110>}-3(Si_1)_{TC}$ | 2.09 | - | 2.13 | 1.19 |
| 4 | 2 | 2 | $(Si_1)_{TC}-(C_2)_{BC'}-(Si_1)_{TC}$ | 1.95 | - | 1.96 | 1.25 |
| | 3 | 1 | $(Si_1)_{TC}-3(C_2)_C$ | 2.04 | - | 2.10 | 1.08 |
| | 4 | 0 | $2(C_2)_{BC'}$ | 1.70 | - | 1.72 | 1.33 |

[a] The GS structure optimized in this study is defined as $(Si_1)_{TC}$ where few of the surrounding Si and C sublattice atoms are displaced. It should be noted that the notation in Ref [16] is different from the one used in this study.

[b] Ref [47] values are adjusted to have the same chemical potential as used in the current study and in Ref [16].

Table 4. The binding energy of G-W-optimized SIA clusters. The binding energies ($E_B$, $E_B'$, and $E_B''$) are defined as the change in the enthalpy when a cluster dissociates into interstitial(s) or sub-clusters. Binding energies are defined in Equation 3.

| $n$ | $n_C$ | $n_{Si}$ | $E_B$ (eV/Int.) | | $E_B'$ (eV/Int.) | | | $E_B''$ (eV/Int.) | | |
|---|---|---|---|---|---|---|---|---|---|---|
| Species unbinding | | | Si | C | Si | Si, C | C | 2Si | Si+C | 2C |
| 1 | 0 | 1 | - | - | - | - | - | - | - | - |
| | 1 | 0 | - | - | - | - | - | - | - | - |
| 2 | 0 | 2 | 1.94 | - | - | - | - | - | - | - |
| | 1 | 1 | 1.08 | 1.08 | - | - | - | - | - | - |
| | 2 | 0 | - | 1.42 | - | - | - | - | - | - |
| 3 | 0 | 3 | 3.18 | - | - | - | - | 1.24 | - | - |
| | 1 | 2 | 1.78 | 0.92 | - | - | - | 0.92 | 1.78 | - |
| | 2 | 1 | 1.27 | 1.61 | - | - | - | - | 1.61 | 1.27 |
| | 3 | 0 | - | 2.06 | - | - | - | - | - | 2.06 |
| 4 | 0 | 4 | 2.5 | - | 3.74 | - | - | 1.8 | - | - |
| | 1 | 3 | 1.9 | 1.58 | 3.68 | 2.82 | - | 0.87 | 0.87 | - |
| | 2 | 2 | 2.31 | 2.14 | 3.58 | 3.92 | 3.06 | 0.82 | 1.42 | 0.82 |
| | 3 | 1 | 0.84 | 1.63 | - | 3.26 | 3.24 | - | - | 0.91 |
| | 4 | 0 | - | 1.88 | - | - | 3.94 | - | - | 1.26 |

Below we first discuss small clusters ($n \leq 4$) optimized with the G-W potential. The goal of this discussion is to evaluate the differences between empirical potentials and DFT results and, more importantly, to validate G-W's capability of providing accurate descriptions for SIA clusters.

### 3.2.1. Comparison between G-W and DFT

As mentioned in Sec. 2.2, our G-W and DFT energies and structures often show some discrepancies. A detailed assessment of the errors in the G-W potential for DFT is beyond the scope of this work but here we point out some of the possible sources of the discrepancy and where the agreement suggests we can most confidently use G-W to understand defect trends in SiC. Some of the differences between G-W and our DFT results may be due differences between our DFT calculations and those used in fitting the G-W potential. For instance, the study by Liao *et al.* [24] showed that the relative stability of the $(Si_1)_{TC}$ and $(Si_2)_{Si, <110>}$ structures is crucially dependent on the supercell size, and more significantly, on the *k*-point mesh. Studies using $\Gamma$-point *k*-point found $(Si_1)_{TC}$ as GS (LDA for Ref.[24,48,49], and GGA for Ref.[50]), while studies with *k*-point mesh denser than $\Gamma$-point identifies $(Si_2)_{Si,<110>}$ as GS (LDA for Ref.[24,51]). It was explained in Ref [24], that calculations with $\Gamma$-point sampling is a source of large errors for electronic structure of $(Si_1)_{TC}$ as the lowest unoccupied molecular orbital is resonant with the conduction band. We note that G-W and DFT calculations with $\Gamma$-point mesh both agree in predicting a $(Si_1)_{TC}$ to be the most stable Si interstitial, whereas DFT calculations with denser k-point mesh (our calculation and Ref. [24,51]) predict a split interstitial $(Si_2)_{Si,<110>}$ as the Si interstitial GS.

Despite these differences between G-W and DFT (hereafter we use DFT to refer to *k*-point converged DFT values, i.e., not those done with $\Gamma$-point *k*-point sampling) calculations, we do find some encouraging agreements. In particular, the $E_D$ values for C clusters with G-W and DFT have energies that agree very well, all to within 0.1 eV/Int. Furthermore, all the C-only GS structures from G-W and DFT are essentially identical. While such good agreement is not found for any other composition, we do note that the GS structures for Si=C clusters are at least qualitatively similar between G-W and DFT. Specifically, both methods predicted interstitial atoms approximately on an extra {111} plane. For instance, the $Si_1C_1$ GS with G-W potential was found to be $(Si_1)_{TC}$-$(C_1)_{TSi}$ where DFT found the GS as $(Si_1C_1)_{TC}$.

In contrast to the C-only and Si=C clusters, the Si-only clusters and most of the non-stoichiometric mixed compositions (not Si- or C-only clusters) for $1 \leq n \leq 4$ showed a more notable disagreement between G-W and DFT structures. For a single Si interstitial, the DFT/G-W energies were 10.47 / 4.93 eV for $(Si_2)_{Si, <110>}$, and 11.05 / 3.36 eV for $(Si_1)_{TC}$, showing different GS structure predicted from the two methods. Also, it was found that the Si SIAs do not form a cluster for Si-only SIAs with G-W, and instead, are found in a group of *n* $(Si_1)_{TC}$ s. Although these defects were grouped together and have $E_D > 0$ (i.e., they are energetically more stable than *n* separated $(Si_1)_{TC}$ s), these grouped $Si_{TC}$ had a nearest-neighbor separation of more than 2.55Å, indicating they are not bonded (the Si-Si interaction range is approximately 2.50 Å with the G-W potential), and likely they only cluster due to strain effects. The grouped $(Si_1)_{TC}$s were also reported in the previous G-W potential study by Watanabe *et al.* [26]. Similarly, in mixed compositions (except for the Si=C compositions), the Si SIAs were always found in $(Si_1)_{TC}$ s, however, for the non-stoichiometric mixed compositions the $(Si_1)_{TC}$ were bonded to C SIAs. In DFT optimizations, the Si-only clusters and mixed compositions for $2 \leq n \leq 4$ did form clusters. Interestingly, the DFT optimized *n*=3, and 4 Si-only clusters included a $Si_C$ antisite (e.g. $2(Si_1)_{BC'} - (C_1)_{BC'} - Si_C$ for *n*=3). The G-W and DFT results differ in predicting the most stable configuration of Si only clusters. As noted above, this difference may have its origin in different *k*-point meshing between our DFT results those used to fit the G-W potential. However, the observations of non-bonded clusters from G-W potential, and the formation of antisites rather than Si-only interstitial clusters from DFT, both suggest that the clustering of Si interstitials is weak.

Given the above observations, it seems that calculations based on the G-W potential can provide useful qualitative predictions of the SIA cluster structures and energies for pure C clusters, and perhaps for Si=C clusters. However, Si-only and non-stoichiometric mixed cluster results should be interpreted with caution.

### 3.2.2. Composition and configuration of SIA clusters

To assess the stability vs. isolated interstitials, in Figure 6 $E_D$ from the G-W potential as a function of the cluster size, *n*. Results are plotted for the three types of compositions (Si-only, C-only, and Si=C). The stability comparison between these three compositions can be divided into two stages: (i) for $n < \sim 10$ the $E_D$ of Si-only and C-only clusters are greater than that of Si=C, and (ii) for $n > \sim 10$, the $E_D$ of Si-only and C-only clusters saturates whereas $E_D$ of Si=C becomes greater than those of Si-only and C-only.

For $n \leq 30$, the Si-only cluster all show anti-clustering (but grouped) behavior of Si SIAs, which was reported for $n \leq 4$ in Sec. 3.2.1. Even for larger values of *n*, it was observed that Si interstitials stay as $Si_{TC}$, with nearest-neighbor distances ranging from 2.55 to 3.32 Å. As noted in Sec. 3.2.1, this distance is larger than the Si-Si interaction range of ~2.5 Å, demonstrating that these groups of $Si_{TC}$s are not covalently bonded, and therefore they have a weak (strain-induced) interaction between them. Despite the bonding being weak, these interstitials are still more stable in a group than *n* separated $Si_{TC}$s.

Next, the configuration of SIA clusters is next analyzed in order to explore the defect cluster shape vs. size relationship. The formation energy of a cluster can be expressed as follows:

Equation 4 : $E_F = E_V + E_S = E_V + \gamma_{INT} A$

where $E_V$ is a volume (bulk) energy of a fictitious cluster with no interfaces, and $E_s$ is an isotropic interfacial energy. Then $E_s$ term can be expressed as $\gamma_{INT} A$, where $\gamma_{INT}$ is an effective interfacial energy and *A* is the interfacial area of the cluster. We note that the effect of this interfacial energy is not obvious. If $\gamma_{INT}$ is independent of shape then clusters are expected to form spheres, as this shape minimized the surface area to volume ratio. However, the surface energy is expected to be highly anisotropic, and it is generally expected that large clusters will minimize their interfacial energy by forming a planar (disk shaped) cluster whose interface is a stacking fault (over the in-plane area) and dislocation (around the plane circumference). SiC has a low stacking fault energy [52-54], consistent with the fact that the material often exhibits stacking faults and twins in CVD-SiC [6,15]. Other shapes besides spherical and planar are certainly possible (e.g., the approximately linear defects observed for SIA clusters in Si), but no such geometries are observed for any size of our GA calculations. Therefore, only spherical, planar, and intermediates between them are included in our closed form formation energy expressions below. In practice, the actual shape of clusters as a function of size have not been previously established, and here we describe this shape for our stable clusters.

To characterize the shape of clusters of different sizes, we use principal component analysis (PCA) on the coordinates of the defect cluster structure. The defected atoms are identified as follows. First we identify atoms in the defected region, including both SIA atoms and lattice atoms. When the displacement of an atom in the defected region is greater than or equal to a cut-off distance from the undefected sublattice site, this atom is labeled as defected. The cut-off distance is set as 0.84 and 0.58 Å, for Si and C atoms, respectively. These values were chosen as 75% of the atomic bond radii in interstitial dumbbells (i.e., half of the dumbbell bond length, where the bond length are 2.24 Å and 1.54 Å for Si-Si and C-C, respectively. These radii are also consistent with the bond-length in a Si-C dumbbell of 1.89Å.); the cut-off radii in turn have been estimated to be 0.84 Å for Si and 0.58 Å for C. Antisites were automatically identified as defected atoms, since the species are different compared to undefected sublattice site. Once all the defected atoms are identified then PCA is performed on the coordinates of the defected atoms. Three principle components $x'$, $y'$, and $z'$ are identified where $x'$ is the axis with the most variance, $y'$ with the second most variance, and $z'$ with the least variance. We then introduce a shape factor ($s$), which is a dimensionless quantity that describes the shape of a cluster in terms of its principle components according to the following expression

Equation 5: $s = 2 \Delta z' / (\Delta x' + \Delta y')$

where $\Delta x' = x'_{max} - x'_{min}$ is the change from the maximum to minimum value along the $x'$ axis of the set of all defect atomic positions, and $\Delta y'$ and $\Delta z'$ are defined analogously to $\Delta x'$. The shape factor can be used to assess whether the cluster shape is more spherical or more planar. The $s$ values for an ideal sphere and a plane (without any severe local displacements) are 1 ($\Delta x' = \Delta y' = \Delta z'$) and 0 ($\Delta z' = 0$), respectively.

In Figure 7, we show a color map of $s$ values for G-W potential optimized for (a) C-only clusters and (b) Si=C composition SIA clusters. The labels on the map represent $s$ value at each cluster size. $E_D$ as a function of cluster size is again plotted as a guide to the defect thermodynamic stability. Both Figure 7 (a) and Figure 7 (b) show qualitatively similar trends in $s$ as a function of $n$. For $n \leq 3$, the $s$ value for both C-only and Si=C are zero as up to three SIA reside in a single plane by definition, without causing any large distortion on nearby lattice atoms. For the $n = 4$ case of Si=C composition, the interstitials somewhat anomalously fit into an essentially perfect plane, but this planar behavior does not occur in the $C_4$ cluster or for any clusters with $4 \leq n \leq 30$. In this range of the $n$, values of $s$ for stable clusters decreased as $n$ increased for both C-only and Si=C compositions. This trend is expected to continue for larger clusters, although such clusters become increasingly computationally challenging to explore with a full GA GS search so this work stopped at 30 interstitial clusters. It was found that the $s$ value ranged from 0.9 to 0.6 for C-only clusters, and from 0.8 to 0.4 for Si=C clusters. It should be noted that theoretically the $s$ value for a perfect interstitial {111} loop would not be 0 for smaller loops but would only approach 0 asymptotically as $n$ goes to infinity. For instance, when we artificially create a perfect interstitial {111} loop of $n = 30$ interstitials, we found the shape factor to be $s = 0.25$. In contrast, the GA algorithm using G-W potentials predicts $s = 0.43$ for $n = 30$ cluster, demonstrating the G-W stable cluster is far from a perfect interstitial {111} loop even at 30 interstitials. We expect that the shape factor will continue to decrease approximately monotonically for increasingly larger clusters. For the Si=C cluster, we expanded the calculation to include just $n = 50$ (which uses a 8×8×8 conventional simulation cell) to demonstrate that the trend in morphology (i.e., the change in $s$) continues for larger $n$. The $s$ value is found for the $n = 50$ cluster predicted by GA is 0.21 whereas the $s$ value for a perfect {111} loop of the same size is found to be 0.18. These observations indicate that there is a slow evolution in the configuration of the most stable clusters, where small clusters (for $n > 4$) are stable in more spherical geometries while larger clusters are stable in a more planar configuration.

## 4. Discussion

In this work, an automated GS search was conducted to identify GSs of SIA clusters in SiC across compositions, using both DFT and potentials. In the first part of this work, we presented findings of the lowest-energy configurations of small interstitial clusters using DFT calculations. Our approach successfully predicted previously reported GSs as well as new GS defect cluster configurations that had not been identified in earlier studies. In the second part of the paper, the G-W empirical potential was utilized to optimize clusters to predict the stability of relatively larger clusters of $n \leq 30$, and we demonstrated both new cluster structures and new trends in the overall shape with an increasing cluster size. Our results suggested that there are compositional and configurational transitions of most stable $I_n$ cluster vs. $n$-SIAs.

Here, we first discuss which clusters of SIAs are stable and how this couples to composition and size. Both DFT and G-W potential results suggest that clusters are stable with respect to decomposition into interstitials and/or other clusters for all compositions, even at very small sizes, except for DFT optimized Si-only clusters. Due to their anomalous behavior, we will discuss DFT Si-only clusters later and we first focus C-only and mixed compositions from the DFT, and Si-only/C-only/Si=C compositions from the G-W potential. In DFT optimizations, the positive values of both $E_D$ (Table 1) and $E_B$ (Table 2) demonstrated the clustering of mono-SIAs is energetically favorable. Moreover, the observation of favorable clustering of sub-clusters ($E_B' > 0$, and $E_B'' > 0$) implies that the growth into larger sizes is favored for small SIA (sub-) clusters, or equivalently, that they will not decompose into subclusters. Similar analyses were performed for the clusters optimized with the G-W potential, which also showed positive values of both $E_D$ (Table 3) and $E_B$ for all three compositions (Si-only, C-only, and Si=C for G-W potential) for cluster with $n \leq 30$. The G-W potential result implies that clustering, and the growth to larger clusters, is thermodynamically favorable, even for the Si-only composition.

With respect to the Si-only clusters, one observation from the DFT result is the inclusion of an antisite defect ($Si_C$) in the cluster of size $n=3$. The formation of an antisite for $Si_3$ composition (($Si_2C_1)_{BC} - Si_C$) indicates that the trimer Si interstitials is unstable, presumably because there is not enough space to have three large Si atoms occupy nearest-neighbor BC sites and forming an antisite is energetically preferred. This observation suggests that antisites might also form for larger values of $n$, where the strain effects are expected to be greater. Indeed, the observation of an antisite in the DFT $n=4$ Si-only structure supports this idea and strongly suggests that formation of an antisite is anticipated for even larger clusters. Another surprising observation from DFT is that the growth of the Si$_3$ cluster is energetically unfavorable, i.e., the reaction Si$_3$+Si$_1$ →Si$_4$ would increase the formation energy. Lack of the driving force for this reaction (reflected in the negative binding energy) may have its origin in the same strain effects that were proposed above to account for the antisite formation in the Si$_3$ cluster. Due to its large size relative to C, placing many Si defects bonded together may lead to a large strain field and therefore to a high-energy cost. Thus, our DFT predict that Si-only SIA clusters are unlikely to grow. Larger Si clusters ($n > 3$) are likely unstable (based on DFT calculations with sizes of up to $n = 4$).

In terms of dissociation energy, both DFT and G-W potential predicted the values of $E_D$ of Si-only clusters to be positive, which means these clusters are stable as compared to $n$ isolated interstitials. While two methods consistently predicted the $E_D > 0$ for Si-only SIAs, the G-W potential results differ in some respect from the DFT results. The Si-only clusters optimized with G-W potential showed a weak clustering tendency for Si by forming only non-bonded Si$_{TC}$ clusters for $n \geq 2$. As discussed in Sec. 3.2.1, the grouped Si$_{TC}$s had separation distances greater than the distance for Si-Si covalent interaction, indicating the Si$_{TC}$s has no (or weak) direct bonding, and likely the grouping is driven by reduction in the strain energy of these defects. Similar conclusions are obtained by da Silva *et al*. [55] from a Monte Carlo study of amorphous SiC with a Tersoff empirical potential. They found in a simulated annealed amorphous SiC that 88% of C atoms were in small clusters whereas 92% of Si atoms were in a single, large, and sparse network. Although the work by da Silva *et al*. cannot be directly compared to our present study as there are many differences in the simulations, the study draws a similar conclusion of C-C bonding is preferred over Si-Si bonding.

To summarize, the Si-only cluster results from DFT and G-W potential suggest a weak and anti-clustering behavior, respectively. Thus, both methods suggest that Si SIAs are unlikely to grow while maintaining their composition. More likely Si-only clusters transition to clusters with a finite C concentration, and for large enough clusters, likely with the Si=C composition. To our knowledge, this is a first evidence to show that the BSDs are unlikely to be rich in Si.

In optimization of relatively larger clusters with $6 \leq n \leq 30$, the cluster stability was investigated with the G-W potential. The analysis above included three compositions: Si-only, C-only and Si=C. However, we only investigate C-only and Si=C in detail, as the previous discussion suggested that larger Si-only clusters ($n > 3$) are likely unstable and dissociate in to smaller Si-only clusters. Mixed compositions other than Si=C are not investigated in these G-W studies. This restriction was necessary to make this study tractable, as exploring all compositions would lead to an enormous number of calculations. However, larger clusters approach a Frank dislocation loop structure, which has stoichiometry Si=C. It is therefore likely that for clusters with mixed composition, the stoichiometric clusters play a dominant role for many conditions. As can be seen from Figure 6, $E_D$ increase as $n$ increases (which implies that the binding energy $E_B$ for an additional interstitial is always positive), predicting that the cluster growth will be favorable for C-only and Si=C clusters at all values of $n$ studied here. The C-only clusters were more stable than Si=C at $n < \sim10$ and the Si=C cluster became the most stable for $n > \sim10$; these calculations are done using isolated C and Si interstitials as reference states. As $n$ increase, planar type clusters are found to be increasingly stable vs. more spherical clusters. In the previous study Watanabe *et al.* [26] investigated the relative cluster stability of $n$ = 100-300 clusters and found the order of decreasing stability to be for compositions: Si=C, Si-only, and C-only. However, Watanabe *et al.* considered only a local optimization of SIA structures initialized in close-packed (111) planes. The compositional dependence of SIA cluster stability in our global optimization confirmed that the Si=C clusters are the most stable for $n \geq 10$. We have also shown that the most stable clusters with $n < 10$ are not Si=C clusters. Furthermore, we also found that the {111} interstitial loop, which is generally assumed to be the stable structure for large interstitial clusters, is not stable for clusters of size $n \leq 50$. As described in Sec. 2, the {111} planar structures were included as candidates for GS structures in the initial GA populations, but our calculations (C-only/Si-only/Si=C, and $n \leq 50$) did not identify the {111} interstitial loop to be the GS. Our shape analysis of clusters suggests that cluster in the more planar {111} interstitial loop structure will be more stable as *n* increase.

With the results from DFT and G-W, we now can make predictions regarding the composition and shape of thermodynamically stable SIA clusters of a given size, *n*. (i) The Si-only cluster ($n>3$) are likely unstable (based on DFT) and are thermodynamically not favorable to grow into larger Si-only clusters. (ii) At small sizes $3 \leq n \leq 10$, SIA clusters are most stable as C-only. For clusters with larger sizes ($n >10$), the G-W results suggest that clusters are more stable in a Si=C, or nearly Si=C, composition. (iii) In terms of the shape of clusters, there is an evolution of stable clusters from more spherical to more planar, starting above 4 interstitial clusters.

It is useful to provide a single functional relationship for how the cluster energies depend on the cluster size, which information is important to provide understanding of how the cluster evolves, and in what shapes the clusters are stable. We focus on the Si=C, as they are most stable for $n >$ 10. The GA calculated formation energies from both DFT and G-W are summarized in Figure 8 (a). Quite large absolute discrepancies can be seen, mostly due to large differences in the energies of the isolate interstitial. Analytically, the formation energy of cluster also can be calculated according to Equation 4. In Sec 3.2.2, we have shown that there is a clear trend in the shape of stable SIA clusters with their size. However, the second term on the right-hand side of Equation 4, $\gamma_{INT} A$, depends on the shape of the cluster. This dependence must be included in the model of formation energy to get a simple expression of formation energy vs. size. Therefore, we first fit the shape factor (*s*) as a function of cluster size to an exponential function, and we will then use that fit in our formation energy model. The fitting yields $s(n) = \exp(-0.030 \cdot n)$ with chi-squared value of 0.002. This fit provides a good representation of the calculated data, as can be seen from the inset of Figure 8 (b).

In the following, we present a general approximate formula for GS energies of Si=C cluster by combining expression for formation energies of spherical and planar clusters, $E^{SPH}$ and $E^{PLN}$, with a weighting for their contribution given by the shape factor as follows

Equation 6: $\quad E_F(n) = s(n) \cdot E^{SPH}(n) + (1 - s(n)) \cdot E^{PLN}(n)$

For the spherical SIA cluster, the $\gamma_{INT}A$ term in Equation 4 can be expressed as $4\pi r^2 \cdot \gamma^{SPH}$, where $r$ is the radius of the sphere and $\gamma^{SPH}$ is the interfacial energy of the cluster per unit area. The radius is related to the number of interstitials through the equation $4/3\,\pi r^3 = n\Omega$, in which $\Omega$ is the volume of an interstitial in our 1728-atom simulation cell. For $\Omega$, we used 3.46 Å$^3$, which is the mean value of the change in volume resulting from insertion of a C or a Si interstitial; these volume changes are 1.4 and 5.49 Å$^3$, respectively. From these relationships, $E_{SPH}$ can be calculated as

Equation 7: $\quad E^{SPH}(n) = E_V^{SPH} + ((4\pi)^{\frac{1}{3}}(3\Omega)^{\frac{2}{3}} \cdot \gamma^{SPH})n^{-\frac{1}{3}}$

where $E_V^{SPH}$ is the volume energy in a spherical cluster. For the planar SIA cluster, we assume the structure to be a {111} Frank dislocation loop [6]. In this case, the formation energy is equal to $\pi r^2 \cdot \gamma^{SF} + 2\pi r \cdot E^{FD}$, where $\gamma^{SF}$ is the stacking fault energy per unit area of the loop, and $E^{FD}$ is the total energy of the Frank dislocation (FD) per unit length of the dislocation [56]. The radius of the loop is related to the number of interstitials through the equation $\pi r^2 = n\rho$, where $\rho = 2.1$ atom/Å$^2$ is the atomic area of an infinite {111} Frank dislocation loop (i.e., no edge effect of the loop) in the {111} projection. For planar SIA cluster $E_F$ can therefore be written:

Equation 8: $\quad E^{PLN}(n) = (\rho\gamma^{SF}) + \left(2(\rho\pi)^{\frac{1}{2}} \cdot E^{FD}\right) \cdot n^{-\frac{1}{2}}$

In S.I. Section C, the $\gamma^{SF}$ and $E^{FD}$ are evaluated explicitly using the G-W potential from separate calculations of stacking faults and Frank dislocations, respectively. We find $\gamma^{SF} = 0.013$ eV/Å$^2$ and $E^{FD} = 1.66$ eV/Å, and further details can be found in S.I. Section C. It should be noted that the dislocation energy has a strain field component that is dependent on the system size, and therefore $E^{FD}$ needs to be evaluated for the relevant system size. Here we used $E^{FD}$ for a $2.62 \times 2.62 \times 2.62$ nm$^3$ cell, which is the simulation cell size used in GA calculations (the $n = 50$ GA calculation was an exception, as this was done in $3.05 \times 3.05 \times 3.05$ nm$^3$ cell to avoid interactions across the cell boundary during the GA optimization. For consistency, we have recalculated the $E_F$ using the GA $n = 50$ structure in a $2.62 \times 2.62 \times 2.62$ nm$^3$ cell and this value is used in Figure 8. The use of the smaller cell introduces a 0.11 eV/Int. change in the energy due to cell size effects, but we expect this to have a minimal impact on the overall $E_F$ fit.

With Equations 6–8 and values above, we fit the unknowns $E_V^{SPH}$ and $\gamma^{SPH}$ to GS energies for $10 \leq n \leq 50$ GA results and obtained $E_V^{SPH} = 1.02 \pm 0.15$ eV and $\gamma^{SPH} = 0.03 \pm 0.01$ eV/Å$^2$. In total, the GS formation energies (eV/Int.) of Si=C cluster can be written as

Equation 9: $E_F(n) = s(n) \cdot \left(1.02 + 0.37n^{-\frac{1}{3}}\right) + (1 - s(n)) \cdot \left(0.03 + 8.49n^{-\frac{1}{2}}\right)$

and the fitted $E_F(n)$ is shown in Figure 8 (b). We exclude results for $n \leq 10$ due to the clusters being so small that they cannot be expected to have a constant interfacial energy. As can be seen from Figure 8 (b), these smaller clusters do in fact have a higher energy than predicted by the constant interfacial energy curve. While one could introduce a size dependent interfacial energy for $n \leq 10$ it seems unlikely to be very accurate as the clusters appear to have non-smoothly varying interfacial energies (see Figure 8 (b)). Furthermore, we note that our results suggest that the stable clusters at these smaller sizes are likely to be C-rich. Because we used a physically motivated functional form for fitting, we believe that this model for $E_F(n)$ can be used to estimate the energy of almost any size Si=C cluster in the given simulation cell size or effective range of the dislocation strain field.

We can express the above fit in a more general form that explicitly includes the effective range of the dislocation strain field by expanding $E^{FD}$ in terms of the core energy and a cylinder radius (a cylinder with an axis along the dislocation line) dependent elasticity term (see S.I. Section C). The modified expression reads

Equation 10: $E_F(n, R) = s(n)\left(1.02 + 0.37n^{-\frac{1}{3}}\right) + (1 - s(n))\left(0.03 + 5.12 \cdot (0.89 + (0.56 \cdot \ln\left(\frac{R}{1.24}\right))n^{-\frac{1}{2}}\right)$

where $R$ is the radius of influence of the dislocation (which is the upper limit of integration of the strain field from the dislocation). Because of the physically motivated functional form used for this fitting, the model in Eq. 10 for $E_F(n)$ can be used to estimate the energy of essentially any size Si=C cluster. The radius of influence $R$ is typically estimated from the dislocation density. For this work, in S.I. Section C, we have discussed how the effective radius $R$ can be approximated from a cubic cell. However, we note that the value of $\gamma^{SF}$ found in this work shows significant errors vs. experiment, suggesting that these energies must be used with some caution (see S.I. Section C). It is worth nothing that the discrepancy on the stacking fault energy may arise from the fact that electronic effects are absent in empirical potential. A single stacking fault layer in 3C-SiC is similar to 4H-SiC locally, and hexagonal 4H has quite different electronic structure compared to 3C-SiC [57]. Moreover, first principle calculations by Iwata *et al*. [58] showed that the SF in 3C-SiC give rise to localized band states in the band gap, and the affect of these states on the stacking fault energy likely are not captured by the G-W empirical potential.

The results in this work can help understand the shapes of small BSDs in SiC. By extrapolating $E_F(n)$ for perfect {111} interstitial loops (Equation 8) and the fitted $E_F(n)$ for the GA optimized clusters (Equation 9), the planar cluster is expected to become stable at $n =$ ~60. This analysis can be found in the S.I. Section D. From the experimental observations, the BSDs are typically smaller than 1.5 nm [11]. The number of atoms in a planar BSDs ($n_{BSD}$) can be approximated by a simple geometric argument, $n_{BSD} = [\pi(\bar{d}/2)^2] / [\pi(\bar{r}_{SiC}/2)^2]$, where $\bar{d}$ is the diameter of BSD, and $\bar{r}_{SiC}$ is the bond length of SiC in bulk (1.89 Å). Assuming they are planar type and equal in Si and C SIAs, a BSD with a typical diameter of 1nm has approximately 120 SIAs (10-300 SIA for diameters of 0.5-1.5 nm, respectively). At this mean size, BSDs are expected to be in a planar shape with Si=C composition, based on the GA results that perfect planar cluster is expected to be stable at $n =$ ~60. The minimum diameter, 0.5 nm, only contains ~10 SIAs if we assume that they are planar cluster. However, we predicted that the planar cluster is not stable at this size. If we assume 0.5 nm BSDs are spherical they will contain ~20 atoms, at which size we predicted the shape of cluster to be an intermediate shape between planar and spherical. So overall we expect the smallest BSDs to be intermediate in spherical and planar character, and contain 10-20 SIAs. This simple assessment can qualitatively guide how the experimentally observed black spot defects are understood with respect to their sizes.

5. **Conclusion**

An automated search (genetic algorithm) was performed across compositions with DFT and empirical potential to optimize SIA clusters in SiC. In the DFT calculation for $n \leq 4$, we have found previously known GS structures and found new GSs for $n = 2$ ($Si_2C_1$), $n = 3$ ($Si_3$, $Si_2C_1$, $Si_1C_2$), and $n = 4$ ($Si_4$, $Si_3C_1$, $Si_2C_2$, $Si_1C_3$) clusters. It was shown that the larger Si clusters ($n>3$) are likely unstable (based on DFT calculations for clusters with size up to $n = 4$) and therefore are unlikely to grow by capturing interstitials or other small clusters. This result suggests that even very small clusters in irradiated SiC are mixed composition or pure C. By combining the calculations from DFT and G-W potential we predicted that the composition of clusters is most stable as C-only for a cluster sizes less than 10 SIAs, and most stable as stoichiometric (Si=C) for cluster sizes greater than 10. Moreover, a transition in the shape of the cluster from more spherical to more planar with size is observed for both Si=C and C-only compositions. The shape of stable cluster is expected to be a planar {111} interstitial loop, for cluster sizes greater than, ~ 60 SIAs. Also, the groundstate formation energies of SIA clusters as a function of a cluster size were presented, which can provide input for rate theory and cluster dynamics models of defect evolution in irradiated SiC.

6. **Acknowledgements**

The authors acknowledge funding from the U.S. Department of Energy, Office of Science, Office of Basic Energy Sciences Mechanical Behavior and Radiation Effects program under Award Number DE-FG02-08ER46493. Calculations were performed using the Extreme Science and Engineering Discovery Environment (XSEDE), which is supported by National Science Foundation grant number OCI-1053575. The authors thank C. Jiang and G. Roma for helpful discussions.

7. **References**

[1] R. J. Price, Nucl. Technol. 35 (1977) 2, 320-336.
[2] Y. Katoh, L. L. Snead, I. Szlufarska, W. J. Weber, Curr. Opin. Solid State Mater. Sci. 16 (2012) 3, 143-152.
[3] Y. Katoh, L. L. Snead, C. H. Henager, T. Nozawa, T. Hinoki, A. Ivekovic, S. Novak, S. M. G. de Vicente, J. Nucl. Mater. 455 (2014) 1-3, 387-397.
[4] A. Mattausch, Ab initio-Theory of Point Defects and Defect Complexes in SiC, Dissertation, Der Friedrich-Alexander-Universität Erlangen-Nürnberg zur Erlangungdes Doktorgrades, 2005.

[5] D. Shrader, S. M. Khalil, T. Gerczak, T. R. Allen, A. J. Heim, I. Szlufarska, D. Morgan, J. Nucl. Mater. 408 (2011) 3, 257-271.
[6] Y. Katoh, N. Hashimoto, S. Kondo, L. L. Snead, A. Kohyama, J. Nucl. Mater. 351 (2006) 1, 228-240.
[7] W. J. Weber, N. Yu, L. M. Wang, J. Nucl. Mater. 253 (1998) 1–3, 53-59.
[8] L. L. Snead, J. C. Hay, J. Nucl. Mater. 273 (1999) 2, 213-220.
[9] A. I. Ryazanov, A. V. Klaptsov, A. Kohyama, H. Kishimoto, J. Nucl. Mater. 307–311, Part 2 (2002) 0, 1107-1111.
[10] L. L. Snead, S. J. Zinkle, D. P. White, J. Nucl. Mater. 340 (2005) 2-3, 187-202.
[11] C. Liu, L. He, Y. Zhai, B. Tyburska-Püschel, P. M. Voyles, K. Sridharan, D. Morgan, I. Szlufarska, Acta Mater. 125 (2017) 377-389.
[12] L. L. Snead, T. Nozawa, Y. Katoh, T.-S. Byun, S. Kondo, D. A. Petti, J. Nucl. Mater. 371 (2007) 1, 329-377.
[13] B. Tyburska-Puschel, Y. Zhai, L. He, C. Liu, A. Boulle, P. M. Voyles, I. Szlufarska, K. Sridharan, J. Nucl. Mater. 476 (2016) 132-139.
[14] L. L. Snead, Y. Katoh, S. Connery, J. Nucl. Mater. 367 (2007) 677-684.
[15] Y. R. Lin, C. S. Ku, C. Y. Ho, W. T. Chuang, S. Kondo, J. J. Kai, J. Nucl. Mater. 459 (2015) 276-283.
[16] F. Gao, W. J. Weber, Nucl. Instrum. Meth. B 191 (2002) 1-4, 504-508.
[17] C. Jiang, D. Morgan, I. Szlufarska, Acta Mater. 62 (2014) 162-172.
[18] A. Mattausch, M. Bockstedte, O. Pankratov, Phys. Rev. B 70 (2004) 23, 235211.
[19] A. Gali, P. Deák, P. Ordejón, N. T. Son, E. Janzén, W. J. Choyke, Phys. Rev. B 68 (2003) 12, 125201.
[20] R. Devanathan, W. J. Weber, T. Diaz de la Rubia, Nucl. Instrum. Meth. B 141 (1998) 1–4, 118-122.
[21] F. Gao, W. J. Weber, Phys. Rev. B 63 (2000) 5, 054101.
[22] G. Lucas, M. Bertolus, L. Pizzagalli, Journal of physics. Condensed matter : an Institute of Physics journal 22 (2010) 3, 035802-035802.
[23] M.-J. Zheng, N. Swaminathan, D. Morgan, I. Szlufarska, Phys. Rev. B 88 (2013) 5, 054105.
[24] T. Liao, G. Roma, J. Y. Wang, Philos. Mag. 89 (2009) 26, 2271-2284.
[25] T. Liao, G. Roma, Nucl. Instrum. Meth. B 327 (2014) 0, 52-58.
[26] Y. Watanabe, K. Morishita, A. Kohyama, J. Nucl. Mater. 417 (2011) 1–3, 1119-1122.
[27] A. Kaczmarowski, S. Yang, I. Szlufarska, D. Morgan, Comput. Mater. Sci. 98 (2015) 0, 234-244.
[28] T. Angsten, T. Mayeshiba, H. Wu, D. Morgan, New. J. Phys. 16 (2014) 1, 015018.
[29] G. Kresse, J. Hafner, Phys. Rev. B 47 (1993) 1, 558.
[30] G. Kresse, J. Hafner, J. Phys.: Condens. Matter 6 (1994) 40, 8245-8257.
[31] G. Kresse, Phys Rev B 54 (1996) 11169.
[32] G. Kresse, J. Furthmüller, Comput. Mater. Sci. 6 (1996) 1, 15-50.
[33] G. Kresse, Phys. Rev. B 59 (1999) 3, 1758.
[34] P. E. Blöchl, Phys. Rev. B 50 (1994) 24, 17953-17979.
[35] J. P. Perdew, K. Burke, M. Ernzerhof, Phys. Rev. Lett. 77 (1996) 18, 3865-3868.
[36] S. Plimpton, J. Comput. Phys. 117 (1995) 1, 1-19.
[37] F. H. Stillinger, T. A. Weber, Phys. Rev. B 31 (1985) 8, 5262-5271.
[38] J. Tersoff, Phys. Rev. B 37 (1988) 12, 6991-7000.
[39] D. W. Brenner, Phys. Rev. B 42 (1990) 15, 9458-9471.

[40] M. Z. Bazant, E. Kaxiras, Phys. Rev. Lett. 77 (1996) 21, 4370-4373.
[41] P. Vashishta, R. K. Kalia, A. Nakano, J. P. Rino, J. Appl. Phys. 101 (2007) 10, 103515.
[42] T. Liang, T.-R. Shan, Y.-T. Cheng, B. D. Devine, M. Noordhoek, Y. Li, Z. Lu, S. R. Phillpot, S. B. Sinnott, Materials Science and Engineering: R: Reports 74 (2013) 9, 255-279.
[43] J. Tersoff, Phys. Rev. B 49 (1994) 23, 16349-16352.
[44] C. Jiang, D. Morgan, I. Szlufarska, Phys. Rev. B 86 (2012) 14, 144118.
[45] A. Gali, N. T. Son, E. Janzen, Phys. Rev. B 73 (2006) 3, 033204.
[46] H. Jiang, L. He, D. Morgan, P. M. Voyles, I. Szlufarska, Phys. Rev. B 94 (2016) 2, 024107.
[47] Y. Watanabe, K. Morishita, A. Kohyama, H. L. Heinisch, F. Gao, Fusion. Sci. Technol. 56 (2009) 1, 328-330.
[48] F. Gao, E. J. Bylaska, W. J. Weber, L. R. Corrales, Phys. Rev. B 64 (2001) 24, 245208.
[49] M. Bockstedte, M. Heid, A. Mattausch, O. Pankratov, Silicon Carbide and Related Materials 2001, Pts 1 and 2, Proceedings 389-3 (2002) 471-476.
[50] G. Lucas, L. Pizzagalli, Nucl. Instrum. Meth. B 255 (2007) 1, 124-129.
[51] O. N. Bedoya-Martinez, G. Roma, Phys. Rev. B 82 (2010) 13, 134115.
[52] K. Maeda, K. Suzuki, S. Fujita, M. Ichihara, S. Hyodo, Philos. Mag. A 57 (1988) 4, 573-592.
[53] Y. Umeno, K. Yagi, H. Nagasawa, Physica Status Solidi B-Basic Solid State Physics 249 (2012) 6, 1229-1234.
[54] M. Jiang, S. M. Peng, H. B. Zhang, C. H. Xu, H. Y. Xiao, F. A. Zhao, Z. J. Liu, X. T. Zu, Scientific Reports 6 (2016) 20669.
[55] C. R. S. da Silva, J. F. Justo, A. Fazzio, Phys. Rev. B 65 (2002) 10, 104108.
[56] D. Hull, D. J. Bacon, *Introduction to Dislocations (Fifth Edition)*. Butterworth-Heinemann: Oxford, 2011.
[57] M. Skowronski, S. Ha, J. Appl. Phys. 99 (2006) 1, 011101.
[58] H. Iwata, U. Lindefelt, S. Öberg, P. R. Briddon, Phys. Rev. B 65 (2001) 3, 033203.

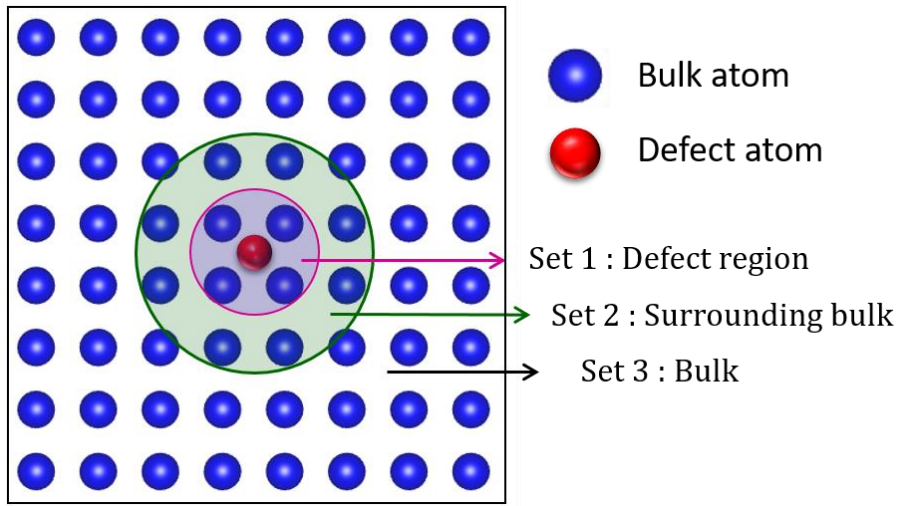

Figure 1. Schematic of regions (sets) in genetic algorithm calculation

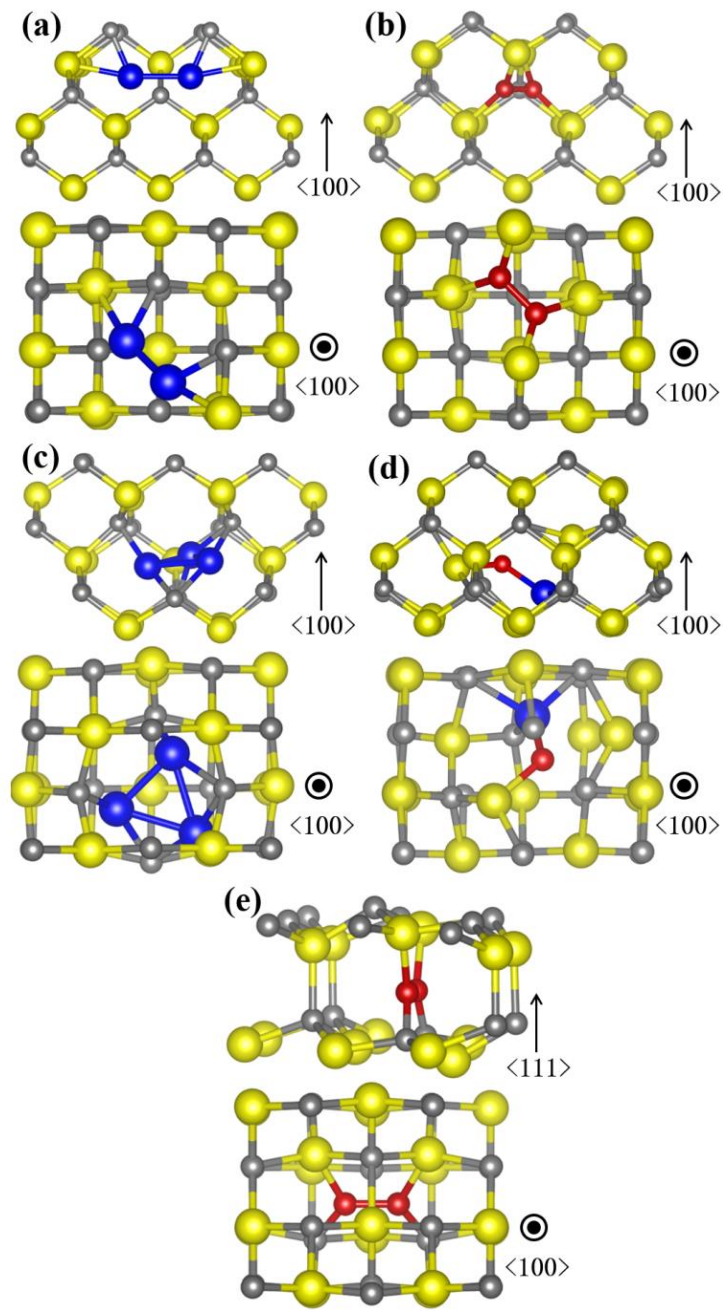

Figure 2. (a-b) Mono- and (c-e) di- SIA cluster GS structures optimized using DFT. Blue/red atoms are Si/C atoms off the original lattice, yellow/gray atoms are Si/C on their expected sublattice sites. (a) mono-Si ($Si_1$) SIA in $(Si_2)_{Si, <110>}$, (b) mono-C ($C_1$) SIA in $(C_2)_{C,<100>}$, (c) di-Si ($Si_2$) SIA in $(Si_3)_{Si, triangle}$, (d) di-Si/C ($Si_1C_1$) SIA in $(Si_1C_1)_{TC}$, and (e) di-C ($C_2$) SIA in $2(C_1)_{BC}$ structure.

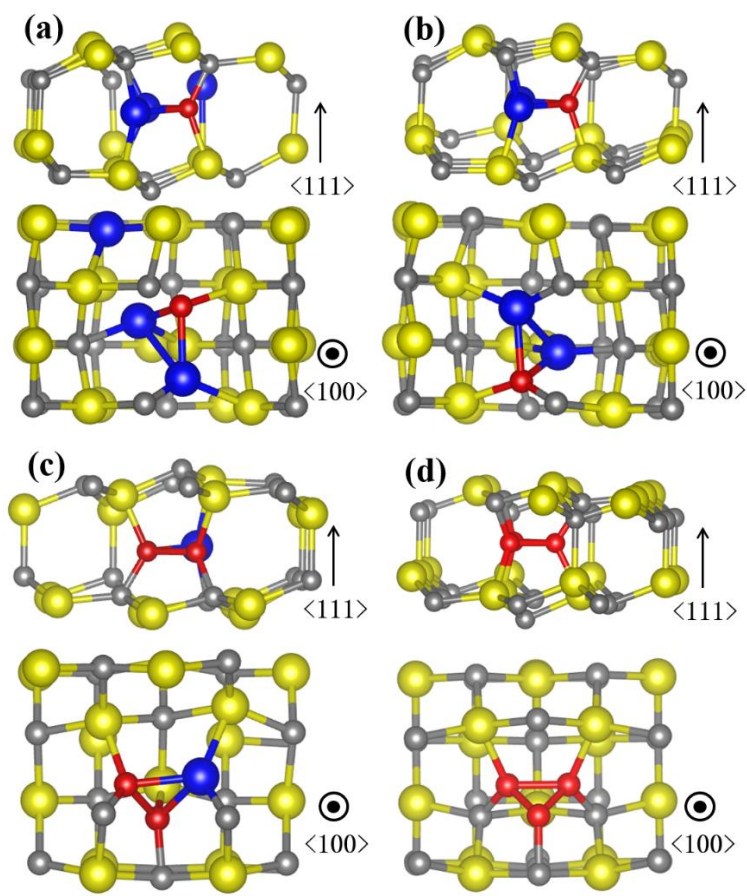

Figure 3. Tri-SIA cluster GS structures optimized using DFT. Blue/red atoms are Si/C atoms off the lattice, yellow/gray atoms are Si/C on the sublattice sites. (a) $Si_3$ SIA in $2(Si_1)_{BC'}-(C_1)_{BC'}-Si_C$, (b) $Si_2C_1$ SIA in $2(Si_1)_{BC'}-(C_1)_{BC'}$, (c) $Si_1C_2$ SIA in $(Si_1)_{BC'}-2(C_1)_{BC'}$, and (d) $C_3$ SIA in $3(C_1)_{BC'}$ structure.

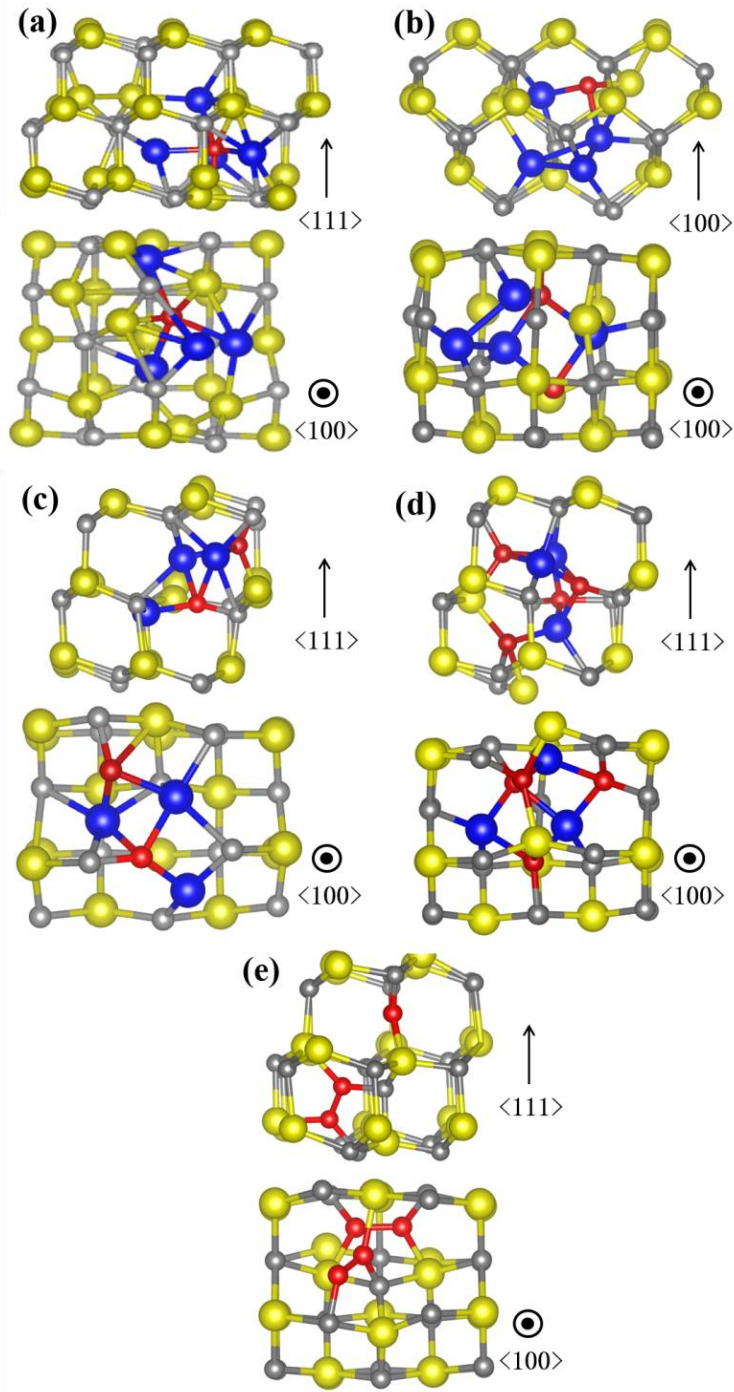

Figure 4. Tetra-SIA cluster GS structures optimized using DFT. Blue/red atoms are Si/C atoms off the lattice, yellow/gray atoms are Si/C on the sublattice sites. (a) $Si_4$ SIA in $(Si_1)_{TC} - 2(Si_1)_{BC'} - (C_1)_{BC'} - Si_C$, (b) $Si_3C_1$ SIA in $(Si_3)_{Si,triangle} - (Si_1C_1)_{TC}$, (c) $Si_2C_2$ SIA in $2(Si_1)_{BC'} - (C_1)_{BC'} - (Si_1C_1)_{Si}$, (d) $Si_1C_3$ SIA in $(Si_1C_1)_{BC'} - 2(Si_1C_1)_{Si}$, and (e) $C_4$ SIA in $2(C_2)_{BC}$ structure.

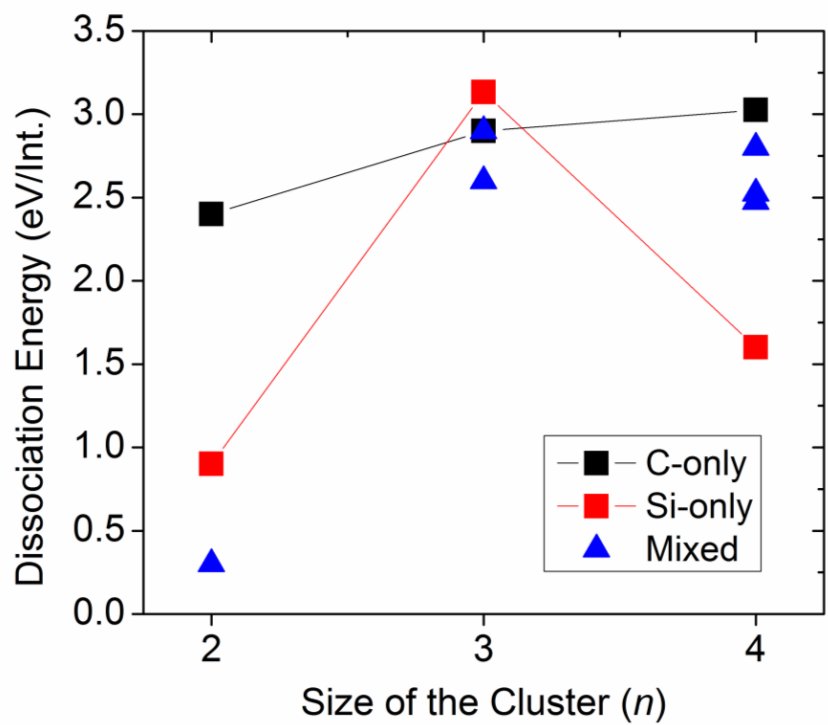

Figure 5. The dissociation energy of DFT optimized SIA clusters as a function of the cluster size for C-only (black), Si-only (red) and mixed (blue) compositions. The solid lines are shown as guides to the eye.

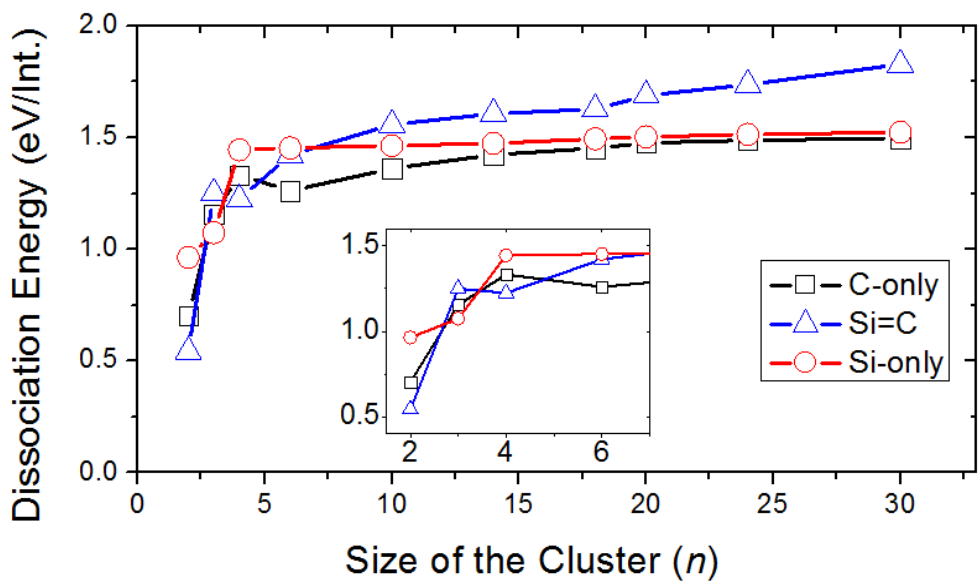

Figure 6. The dissociation energy of G-W optimized SIA clusters as a function of the cluster size for C-only (black), Si-only (red) and Si=C (blue) compositions.

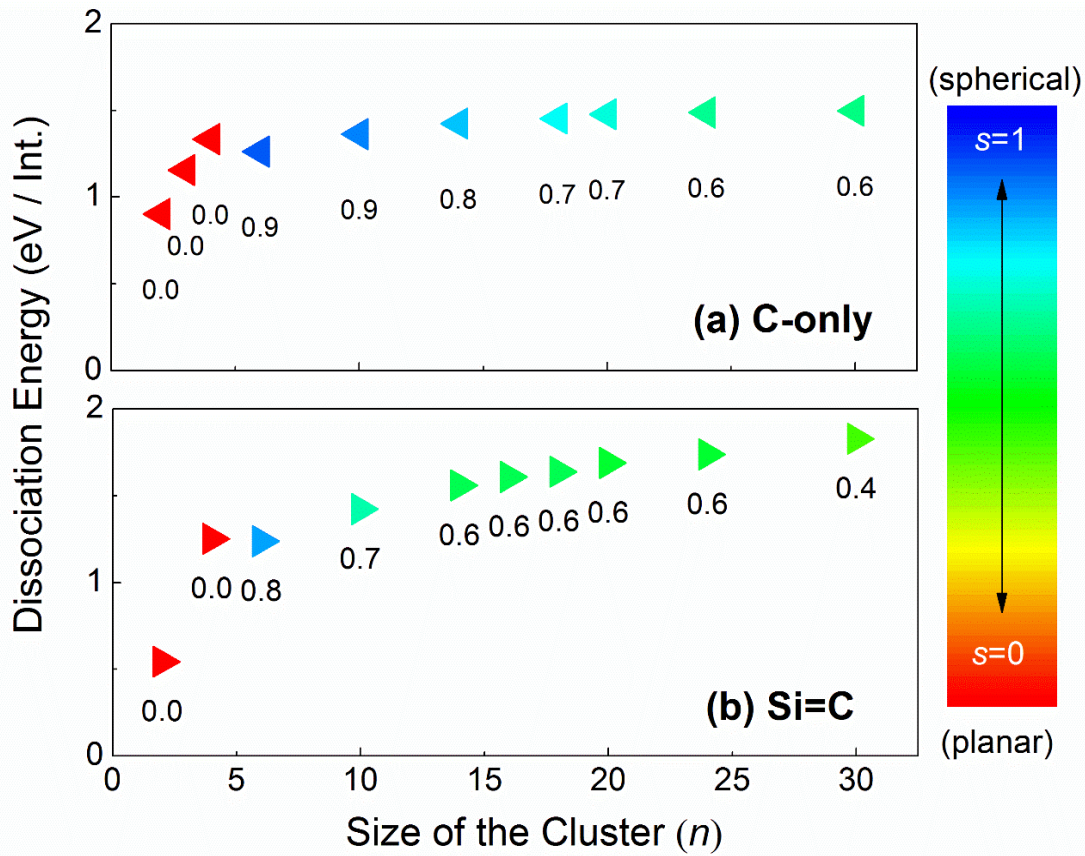

Figure 7. The dissociation of G-W optimized SIA clusters are plotted as a function of cluster size (data equivalent to Figure 6) for (a) C-only and (b) Si=C composition, with a color map showing the shape factor ($s$). The $s$ values at each size are also labeled. If all the SIAs in the cluster can be defined on a single plane, then $s = 0$, and if they are in an ideal sphere then $s=1$.

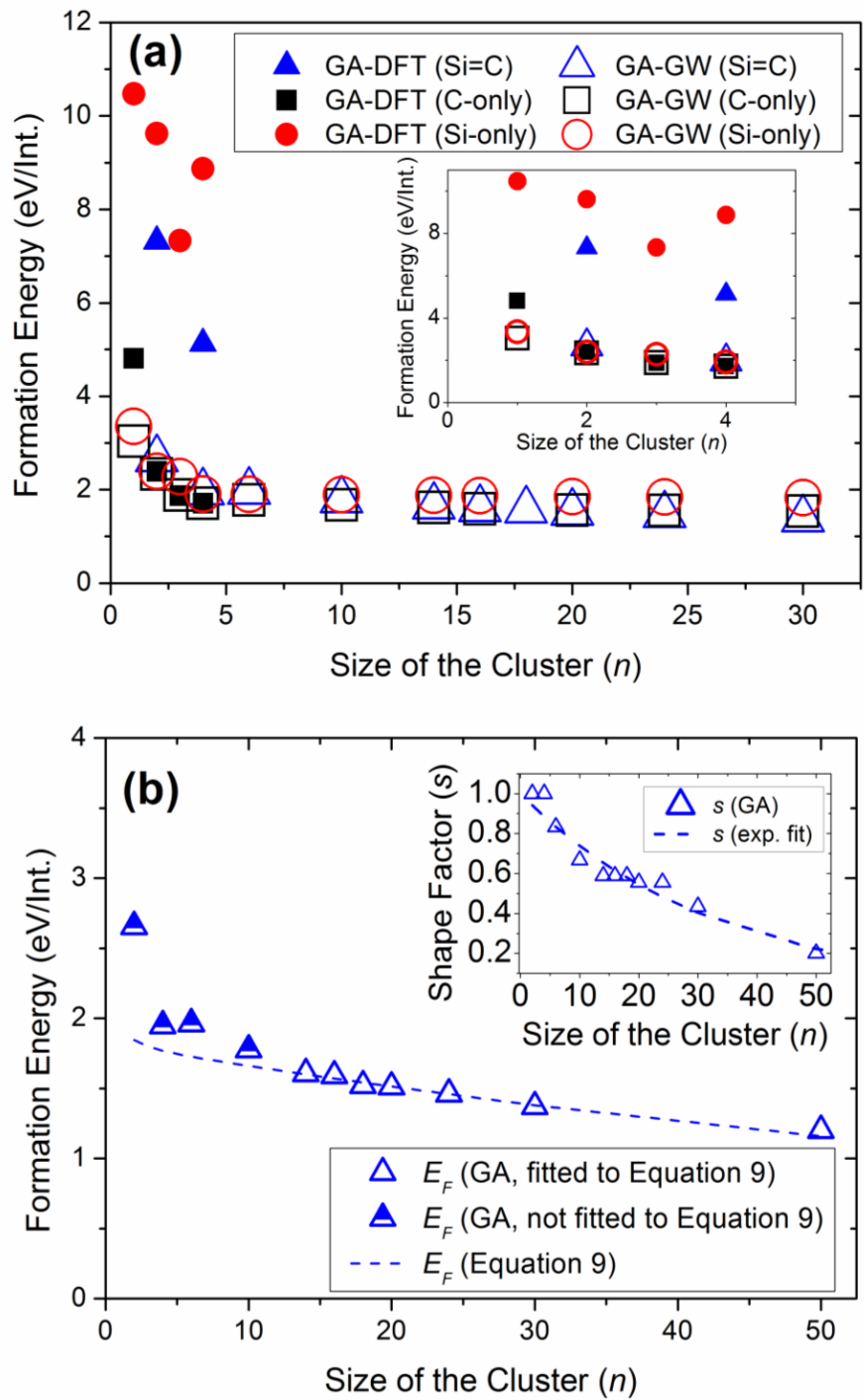

Figure 8. (a) Comparison of formation energies from both DFT (solid symbols), and G-W potential (open symbols) calculations. C-only, Si-only, and Si=C compositions are shown in black, red and blue colors, respectively. The inset shows formation energies for $n \leq 4$. (b) A fitted formation energy as a function of size of the cluster with Equation 9 for GA-G-W with the Si=C composition

(blue dash). The GA results of $n > 10$ (open symbols) are used to fit $E_F$, whereas $n \leq 10$ results (half-filled symbols) are excluded from the fitting. The inset shows the shape factor ($s$) for GA-G-W optimized clusters of Si=C composition (open symbols), and the exponential fit of $s$ (blue dash).